\documentclass[footinbib,a4paper,aps,pra,superscriptaddress,reprint,twocolumn,preprintnumbers,amsmath,amssymb,nobalancelastpage,10pt]{revtex4-1}
\usepackage{amsmath}
\usepackage{bbold}
\usepackage{verbatim}
\usepackage{graphicx}
\usepackage{color}
\usepackage{tikz}
\usepackage{subfigure}
\usepackage{float}
\usepackage{mathptmx}
\usepackage{array}
\usepackage{bm}
\usepackage{graphicx}
\usepackage{braket}
\usepackage[colorlinks=true,citecolor=blue,linkcolor=red]{hyperref}

\usepackage{xcolor}

\begin{document}

\title{Generation of hybrid maximally entangled states in a one-dimensional quantum walk}

\author{Aikaterini Gratsea}\email{gratsea.katerina@gmail.com}
\affiliation{ICFO-Institut de Ci\`encies Fot\`oniques, The Barcelona Institute of Science and Technology, Av. Carl Friedrich Gauss 3, 08860 Barcelona, Spain}
\author{Maciej Lewenstein}
\affiliation{ICFO-Institut de Ci\`encies Fot\`oniques, The Barcelona Institute of Science and Technology, Av. Carl Friedrich Gauss 3, 08860 Barcelona, Spain}
\affiliation{ICREA-Instituci\'o Catalana de Recerca i Estudis Avan\c cats, Lluis Companys 23, 08010 Barcelona, Spain}

\author{Alexandre Dauphin}\email{alexandre.dauphin@icfo.eu}
\affiliation{ICFO-Institut de Ci\`encies Fot\`oniques, The Barcelona Institute of Science and Technology, Av. Carl Friedrich Gauss 3, 08860 Barcelona, Spain}

\begin{abstract}
We study the generation of hybrid entanglement in a one-dimensional quantum walk. In particular, we explore the preparation of maximally entangled states between position and spin degrees of freedom. We address it as an optimization problem, where the cost function is the Schmidt norm. We then benchmark the algorithm and compare the generation of entanglement between the Hadamard quantum walk, the random quantum walk and the optimal quantum walk. Finally, we discuss an experimental scheme with a photonic quantum walk in the orbital angular momentum of light.  The experimental measurement of entanglement can be achieved with quantum state tomography. 
\end{abstract}

\maketitle

\section{\textbf{Introduction}} 

The notion of entanglement plays a central role in quantum information theory~\cite{Horodecki2009} and is a key concept for quantum technologies, as for example quantum teleportation~\cite{Braustein2015}, quantum key distribution~\cite{Scarani2009}, Bell inequalities and non locality~\cite{Rosset_2014, Brunner_2014, Brunner2014}. In this context, the preparation of entangled states is of high importance for applications of quantum technologies.  

Although entanglement is generally studied between different particles, it can also be defined between different internal degrees of freedom of the particles and such entanglement is called hybrid entanglement \cite{Zeilinger2017, Can2005, Aiello_2015, Karimi1172}.  Among others, the experimental preparation of single-particle hybrid entanglement has attracted great interest~\cite{Karimi}. For a single photon, one can study the hybrid entanglement between degrees of freedom such as the polarization, orbital angular momentum, time-bin energy or spatial mode~\cite{Sciarrino2018, Karimi2017}. Although these concepts are at the single particle level, they have applications in metrology~\cite{Ying2018, Zhang:18}, quantum teleportation~\cite{Wang2015}, etc. Furthermore, the possibility of transferring entanglement between particles into hybrid entanglement and vice-versa has recently been realized experimentally~\cite{Vitelli2013}. 

Generally, maximally entangled states have proven to be more useful than weak entangled states~\cite{Zeilinger2018}. Also, the restriction to two dimensional systems is only technological: different schemes for the preparation of high dimensional states have been proposed~\cite{Zeilinger2018} allowing for more information to be encoded in a single particle level. This paves the way for implementations in the field of quantum information and computation. The preparation of high-dimensional maximally entangled states are therefore of great importance: such states could be used, along with quantum gates, for photonic quantum information technologies.  Quantum C-NOT gates for single-photon two qubit quantum logic using polarization and orbital angular momentum have already been proposed~\cite{Deng2007} and realized experimentally~\cite{Canabarro2018}. The C-NOT gates along with single qubit gates allow the implementation of universal quantum computation~\cite{NC}. Also, these states are promising for communication in future quantum networks, since transmission of high-dimensional orbital angular momentum has proved to be possible~\cite{Cozzolino2018, Cozzolino2019}.

One of the platforms to generate hybrid entanglement are the discrete time quantum walks. A discrete quantum walk consists in the repeated application of unitary operators, typically a shift operator and a coin operator. Quantum walks have already been realized experimentally~\cite{Wang_book} in trapped atoms~\cite{Widera2009} and ions~\cite{Roos, Schmitz}, optical lattices~\cite{Meinert} and photonic platforms~\cite{Broome2010, Schreiber2010, Cardano2015}. We here focus on the realization of a photonic architecture and more specifically on single photon implementation. Over the last decades, photonic quantum information has greatly advanced~\cite{Sciarrino2018}, and already offers technological applications, such as  quantum computation~\cite{ Walther2014, Zeilinger2005}, simulation of Floquet topological insulators~\cite{Cardano2015, Massignan2017} and generation of Schr\"{o}dinger cat states~\cite{Zhang2016, Catstates2, Sciarrino2019}. Generally in a quantum walk, the unitary evolution remains the same throughout the walk~\cite{Sciarrino2018}. Recently, there has been an increasing interest in non periodic (in time) Quantum walks. For example, disordered quantum walks generate maximally entangled states in the asymptotic limit~\cite{Rigolin2013, Rigolin2014}. A delocalized initial condition in quantum walks favor the generation of maximal entanglement~\cite{Abal2006, Orthey2017}. Finally, these non periodic quantum walks can also be used for state preparation~\cite{Sciarrino2019}. 

In this work, we explore the preparation of maximally entangled states between the polarization and orbital angular momentum of a single photon with a non periodic quantum walk.   We investigate the preparation of high-dimensional maximally entangled states between the polarization and orbital angular momentum of a single photon with a quantum walk having a fixed number of steps. Importantly, this could resolve the experimental challenges of generating these states, since the fixed number of steps covers for the need of the asymptotic limit~\cite{Wang:18}.

The paper is structured as follows. Section II briefly reviews quantum walks and hybrid entanglement: We tackle the generation of entanglement with a quantum walk. In particular, we study the generation of maximally entangled states and address it as an optimization problem, where the cost function to maximize is the Schmidt norm. Section III presents the numerical benchmarks for the generation of maximally entangled states. Section IV discusses a realistic photonic scheme for the realization and measurement of such quantum walks. Finally, Section V contains the conclusions and outlook. 

\section{Theoretical background}

\subsection{One dimensional quantum walk}

\begin{figure}
\centering
\includegraphics[scale=0.5]{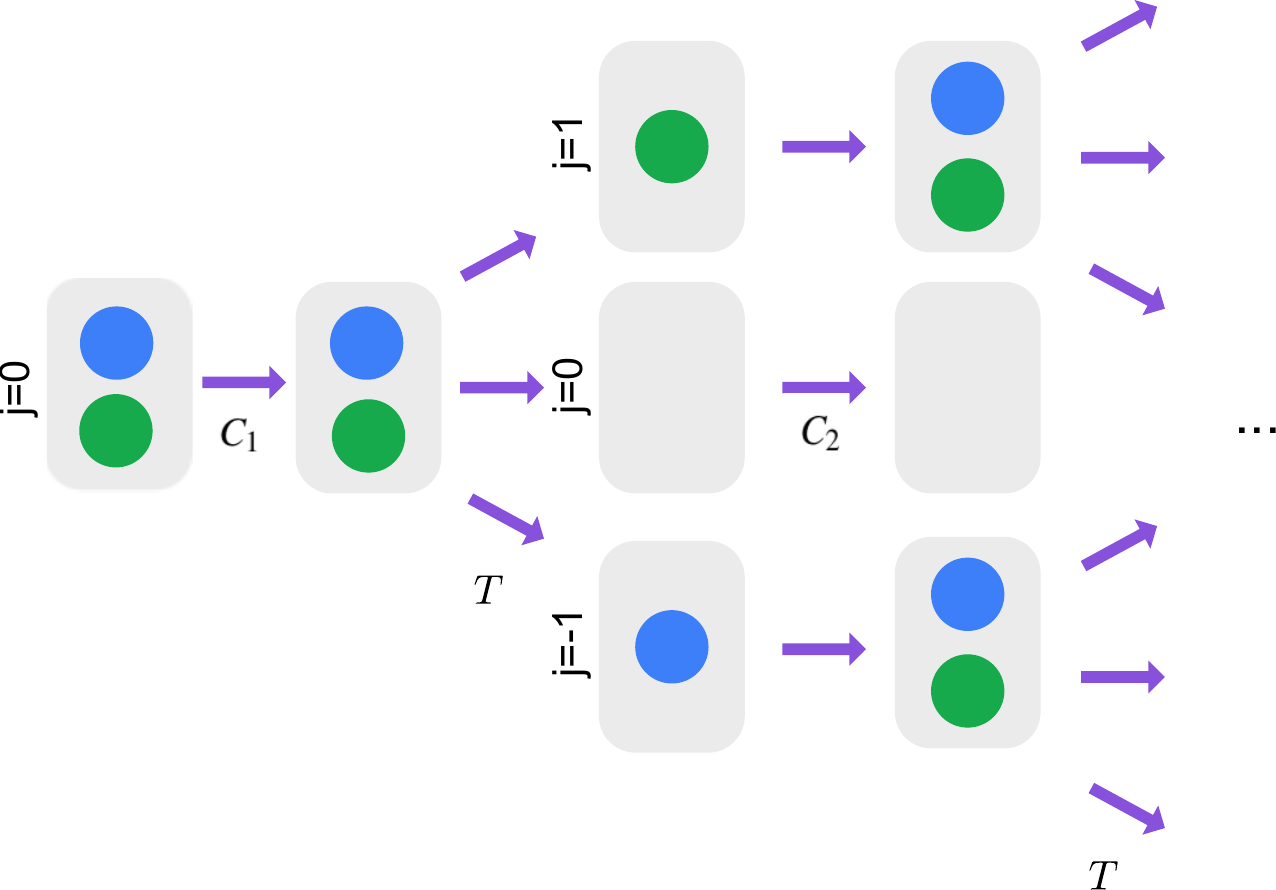} 
\caption {In a quantum walk, the walker is initially localized in the site $m=0$ with an arbitrary superposition over the spin states $L$ (blue) and $R$ (green). Each step $i$ consists in the successive application of a translation $T$ and a rotation $C_i$.}
\label{fig1}
\end{figure}

The quantum walk takes place in a one-dimensional lattice associated to the Hilbert space $H=H_\text{x}\otimes H_\text{C} $, where $H_\text{x}$ stands for the position on the lattice and $H_\text{C}$ for the spin degree of freedom. Initially, the walker is localized on a single site and has an arbitrary superposition of the coin states: $ \ket{\psi_0} =  \sum_\sigma \alpha_{0,\sigma}(0)  \ket{0,\sigma}$. The quantum walk, sketched in Fig.~\ref{fig1}, consists in the consecutive application of a translation operator $T$ and an on-site rotation operator $C_i$, changing at each time step $i$. The translation operator $T$ displaces the walker in different directions, depending  on the coin state

\begin{equation} 
\label{uniS}
T = \sum_m (  \ket{m-1,R}\bra{m,L} +   \ket{m+1,L} \bra{m,R}. 
\end{equation}

The coin operator $C$ rotates the spin degrees of freedom in the following way

\begin{align} 
C_1 =\mathbb{1}_x \otimes \begin{bmatrix}
e^{i\xi_i }cos(\theta_i) & e^{i\zeta_i}sin(\theta_i) \\
e^{-i\zeta_i}sin(\theta_i) & -e^{-i\xi_i}cos(\theta_i)
\end{bmatrix} 
\end{align}  

where $\xi_i,\zeta_i \in [0, 2 \pi]$ and $ \theta_i \in [0, \pi/2]$ are the parameters of the SU(2) rotation~\cite{Chandrashekar2008}. After $N$ steps, the final state reads

\begin{equation} \label{final_state}
\ket{\psi_N} (w)= \prod_{i=1}^N T \, C_i \ket{\psi_0}=\sum_{j,\sigma} \alpha_{j \, \sigma}(N) \ket{j,\sigma}, 
\end{equation}

and depends on the parameters of the rotations  at each step $w=\{\xi_1,\zeta_1,\theta_1,...,\xi_N,\zeta_N,\theta_N\}$.

\subsection{Hybrid entanglement between position and spin degrees of freedom}

\begin{figure} 
\centering
\includegraphics[scale=0.6]{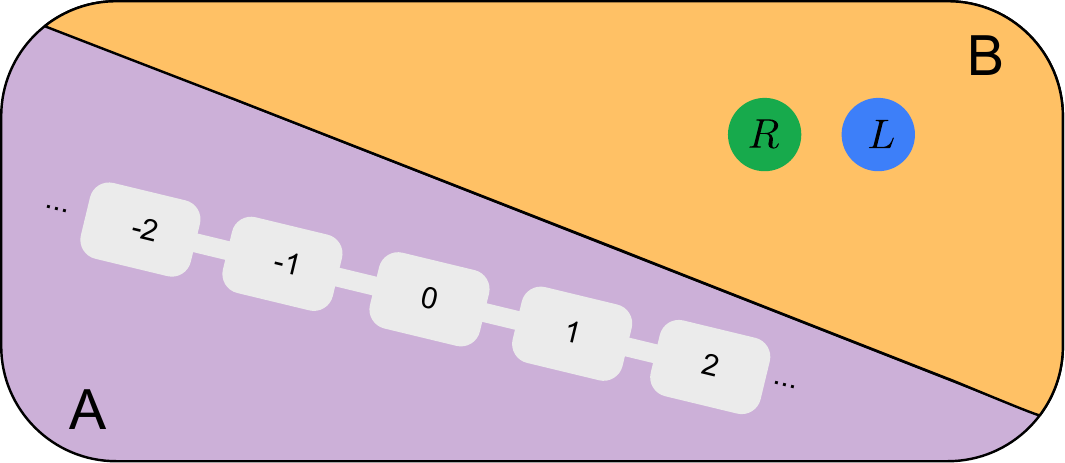} 
\caption {The system is divided to two subsystems: the position ($A$, purple) and the spin  ($B$, orange) degrees of freedom.}
\label{fig2}
\end{figure}

We focus on the hybrid entanglement between position and spin degrees of freedom of a single photon. There are many measurements related to entanglement and we restrict ourselves to the Schmidt norm. Although the system is pure, i.e. $\rho(N)=\ket{\psi_N}\bra{\psi_N}$, we can still define a bipartition between the position and spin degrees of freedom of the photon~(See Fig.~\ref{fig2}). One can perform a singular value decomposition of the matrix $\alpha_{j,\sigma}$ and find its Schmidt coefficients $\lambda_k>0$, where $k\leq \text{min}(d_A,d_B)$ is written in terms of the dimensions $d_A$ and $d_B$ of the Hilbert spaces of the subsystems position ($A$, purple) and spin ($B$, orange). The Schmidt norm is defined as~\cite{Reuvers2018}

\begin{equation} \label{Schmidtnorm}
\parallel \psi \parallel_{p,k} = \left(  \sum_{i=1}^{k} (\lambda_i ^ {\Psi}) ^ {p} \right) ^ {1/p}.
\end{equation}

Throughout the rest of the work, we set $p = 1$. For our system,  $k=\text{min}(d_A,d_B)=2$ and the maximum value of the Schmidt norm for a maximally entangled state is therefore equal to $ \sqrt{2} $, since  $ \lambda_i ^ {\Psi} = 1/\sqrt{k} = 1/\sqrt{2}$. Furthermore, the Schmidt norm can be computed analytically by using the relation between the Schmidt coefficients and the eigenvalues of the reduced density matrix. Indeed, one can write the reduced density matrix as~\cite{Rigolin2013, Zeng2017}

\begin{equation}
\rho_C(N)=\text{Tr}_x(\rho(N))=\frac{1}{2} \, \mathbb{1}+\mathbf{n}\cdot \boldsymbol{\sigma},
\end{equation}

where the trace is taken over the position degree of freedom and 

\begin{equation} \label{n_vector}
\mathbf{n}=(\text{Re}(\sum_i \alpha_{i R}^* \alpha_{iL}),\text{Im}(\sum_i \alpha_{i R}^* \alpha_{iL}),\frac{1}{2}\sum_i\vert\alpha_{i L}\vert^2-\vert\alpha_{i R}\vert^2)
\end{equation}

multiplies the  Pauli vector $\boldsymbol{\sigma}$. The density matrix can then be diagonalized analytically and its eigenvalues  are given by $E_\pm=1/2\pm\vert \mathbf{n}\vert$. Using the identity $E_\pm=\lambda_\pm^2$, one can therefore write the Schmidt norm as

\begin{equation} \label{Schmidt_norm}
S(w)=\parallel \psi \parallel_{1,2}=\sqrt{E_-}+\sqrt{E_+}.
\end{equation}

\subsection{Generation of highly entangled states with a quantum walk}

We want to generate hybrid entanglement and in particular we want to maximize the Schmidt norm $S(w)$ after $N$ time steps. The problem can be reformulated as an optimization problem where the cost function to maximize with respect to the set of parameters $w$ is the Schmidt norm after $N$ steps. In particular, the set of maximal solutions should fulfill $\partial_{w}S=0$, which can be written explicitly as

\begin{equation}
\partial_{w}S=\frac{1}{\sqrt{1-4\vert \mathbf{n}\vert^2}}\partial_w\vert\mathbf{n}\vert (\sqrt{E_-}-\sqrt{E_+})=0.
\end{equation}
 
This optimization problem can directly be related to the minimization of $\vert \mathbf{n}\vert$ or its derivative $\partial_w\vert \mathbf{n}\vert$. This problem has clearly not a unique solution. For example, any Bell states of the form $\frac{1}{\sqrt{2}}(\vert -i,L\rangle + \vert i, R\rangle) $ maximizes the entropy. Such states can be readily realized experimentally~\cite{Karimi}. Furthermore, there is no guarantee of convexity, which makes possible the existence of local minima. In the next section, we will maximize the Schmidt norm numerically and look at the different solutions. In particular, we will focus on the delocalized solutions.

\section{Numerical benchmark}

\begin{figure}
\includegraphics[scale=1]{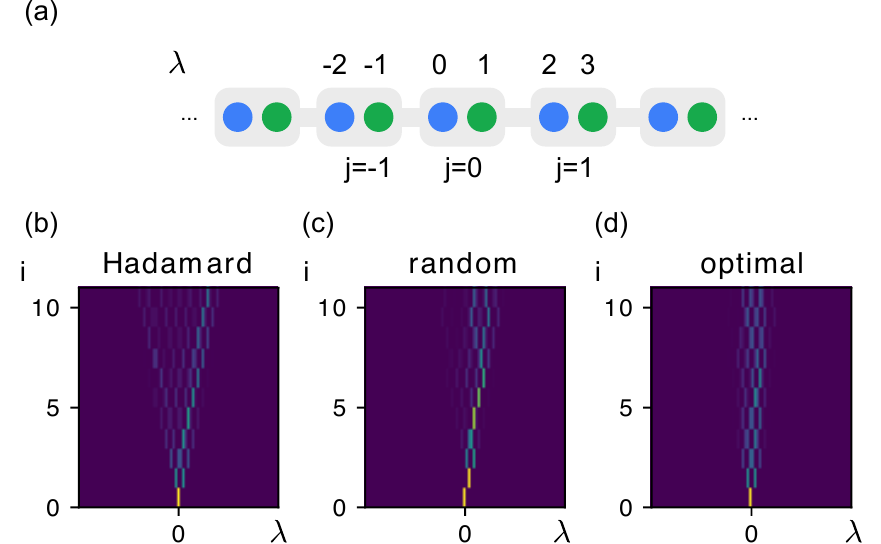} 
\caption{\textbf{(a)} The wave function is plotted in terms the index $\lambda$ which labels both the position and the spin degrees of freedom. \textbf{(b,c,d)} Time evolution of  a quantum walk of ten steps with three different choices on the coin operator: the  Hadamard coin (b) and two others non periodic in time. the coin operator is chosen randomly (c) in the first case and in the second case is chosen with an optimal  set of parameters $w$ (d).}
\label{fig3}
\end{figure}

We numerically maximize the Schmidt norm, starting from an arbitrary initial state. In practice, we use a basin-hopping algorithm~\cite{wales_1997} to minimize the cost function $-S$. The algorithm starts from a random initialization of the parameters of the cost function and performs a local minimization. Then it completes the following cycle: a local perturbation on the coordinates is applied and the algorithm performs a local minimization. The new minimum is accepted or not following the Metropolis rule of standard Monte-Carlo. This algorithm is particularly well suited from problems with local minima as the algorithm allows one to escape from local minima, which is the case in our problem.

\begin{figure}
\centering
\includegraphics[scale=1]{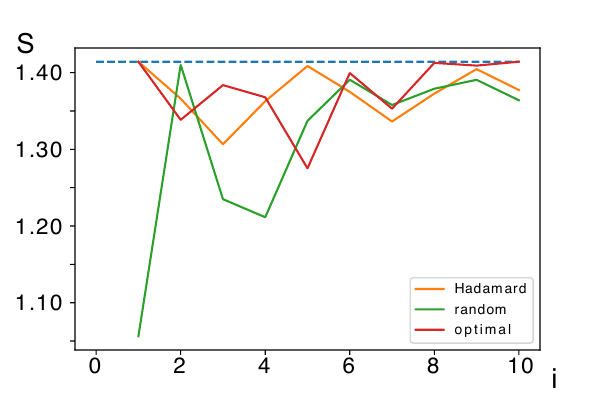} 
\caption{Schmidt norm $S$ during a quantum walk of $N=10$ steps (see Fig.~\ref{fig3} for the density).While the Hadamard and the random quantum walk are highly fluctuating and not converging to the maximal value of the Schmidt norm, the quantum walk with an optimal  set of parameters $w$ generates a highly entangled state.}
\label{fig4}
\end{figure}

We first start to compare the performance of the minimization method with respect to the Hadamard walk and a random walk for $N=10$ steps. We consider an initial state with spin $\vert L \rangle$. Figure~\ref{fig3} shows the density $\vert \psi_i \vert^2$ at different time steps $i$. The spin and position degrees of freedom are labelled by the parameter $\lambda$, as shown in Fig.~\ref{fig3}(a). Figures~\ref{fig3}(b), (c) and (d) shows the Hadamard quantum walk, random quantum walk for one arbitrary disorder realization and a quantum walk with an optimal  set of parameters $w$. The Schmidt norms are shown during the walk in Fig.~\ref{fig4}. On the one hand, both the Hadamard and the random quantum walks do not succeed to reach a maximally entangled state after $N=10$ steps as expected. On the other hand, the quantum walk with an optimal  set of parameters $w$ reaches a state with a very high Schmidt norm after $N=10$ steps.

In order to systematically investigate the performance of the algorithm, we run it with $10 \, 000$ random initial states uniformly chosen over the Bloch sphere. Specifically, we choose the initial states to be

\begin{equation}
\vert \psi \rangle =  cos(\theta/2) \vert 0,L \rangle + e^{i\phi}sin(\theta/2) \vert 0,R \rangle
\end{equation}

where $\theta\in [0,\pi],$ and $\phi \in [0,2\pi]$ are sampled from an uniform distribution. Generally, delocalized initial conditions favor the generation of maximal entanglement~\cite{Abal2006, Orthey2017}.

\begin{figure} 
\centering
\includegraphics[scale=0.55]{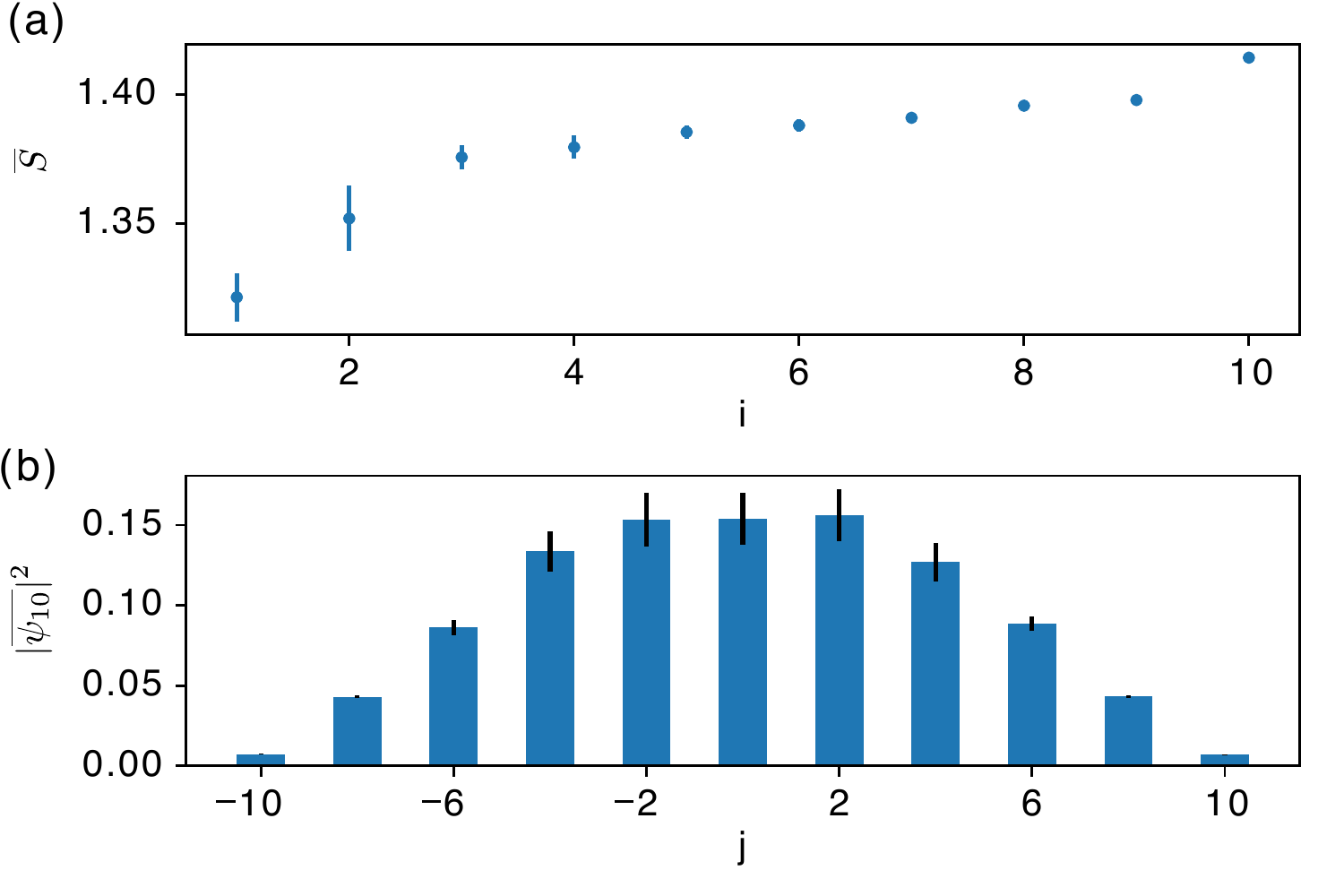}
\caption{\textbf{(a)} Average Schmidt norm during the walk \textbf{(b)} Average amplitude distribution of the final state. The average has been performed over $1302$ out of $10\, 000$ iterations. These states have at least the population at one of the outer sites $\vert-10\rangle$ or $ \vert10\rangle$ is larger than $10^{-4}$. By doing so, we disregard Bell-states or maximally entangled states not spread over all the sites.}
\label{results}
\end{figure}

For all the initial states the algorithm reaches a maximally entangled state. We are particularly interested in high-dimensional states that exploit the whole Hilbert space. Thus, we study the statistics of the final states ignoring the Bell states, which have already been studied~\cite{Karimi}, and states that do not exploit the whole Hilbert space of the position. The selection criteria is chosen such that the population at one of the outer sites $\vert -10\rangle$ or $\vert 10\rangle$ to be larger than $10^{-4}$. Therefore, we have $1302$ out of $10 \, 000$ iterations satisfying the requirement. Figure~\ref{results}(a) shows the average Schmidt norm $\bar{S}$ after each step of the walk: the average entanglement after each step smoothly increases until it reaches the maximum value. Figure~\ref{results}(b) shows the average population of the final states of the walk that exploit the whole Hilbert position space. Interestingly, the definition of the shift operator imposes restrictions to the allowed sites: either odd or even sites survive. 

\begin{figure}
\centering
\includegraphics[scale=1]{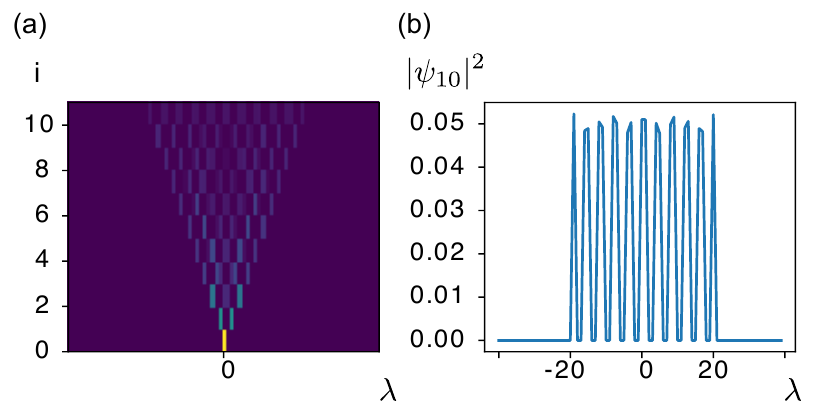}
\caption{ \textbf{(a)} Quantum walk with parameters $w$ maximizing $S+\beta I$, where $\beta=0.1$. The figure shows the density at each step of the quantum walk. \textbf{(b)} Density of the final state $\vert \psi_{10}\vert^2$.}
\label{spread}
\end{figure}

To maximize the spreading of the wave-function in position, we propose to change the maximization procedure by maximizing $S+\beta I$, where $I$ is the inverse participation ratio defined as $I=\Sigma_{i,\sigma}1/\vert\alpha_{i\sigma}\vert^4$ and $\beta$ is an arbitrary positive constant. The addition of this second term will help to spread the wave-function over all the sites. Figure~\ref{spread} shows the results of the optimization algorithm for $\beta=0.1$. Due to the extra constraint on the inverse participation ratio, the wave function at end of the quantum walk is spread on all the even sites, as shown in Fig.~\ref{spread}(b).

\section{Experimental scheme}

\paragraph{Preparation of the state} The aforementioned quantum walk can be implemented in the orbital angular momentum (OAM) and polarization of light~\cite{Cardano2015, Cardano2016, Massignan2017}. The position can be mapped to the OAM as a synthetic dimension and the spins can be mapped to the polarization of light. Alternatively, the position can also be mapped to the momentum of light~\cite{Derrico2018}. The shift operator is realized by a q-plate~\cite{Cardano2015, Cardano2016, Massignan2017} (respectively a g-plate~\cite{Derrico2018}) and the coin operator is realized by a wave plate. 

\begin{figure}
\centering
\includegraphics[scale=1.1]{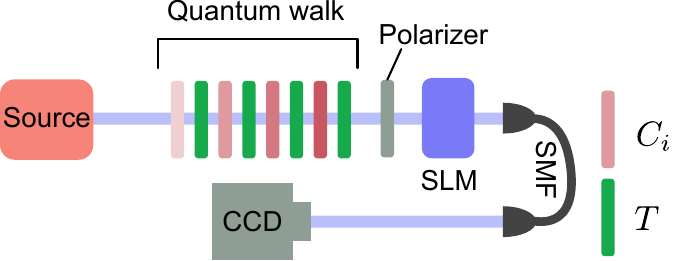}
\caption{Experimental scheme for measuring the amplitudes of the high-dimensional states of a single photon. The state $\bra{\psi_N}$ is generated by a quantum walk , which consists in $N$ sets of q-plates (shift operator) and wave plates (rotations). After the walk, the state is roated and projected on the $L$ polarization. Then, the measurement of the intensity for each OAM state is performed with the help of a spatial light modulator and a single mode fiber, which allows the selection of a specific orbital angular momentum mode. This procedure allows one to extract both the amplitudes and the phases on each site.}
\label{exp_prop2}
\end{figure}

\paragraph{Measurement of the state} The final state $\ket{\psi_N}$ is a superposition of different OAM sites $\vert j\rangle$ and spin degrees of freedom $\vert\sigma\rangle$ \eqref{final_state}. The proposed experimental scheme is sketched in Fig.~\ref{exp_prop2}.  After the quantum walk, the state then reads

\begin{equation} \label{psi}
\ket{\psi_N} =\sum_{j,\sigma} \alpha_{j \, \sigma}(N) \ket{j,\sigma}.
\end{equation}

The full state tomography is a difficult problem as it would require to measure both the amplitude on each site and the phases. Nevertheless, in our case, as it can be seen in Eq.~\eqref{n_vector}, we only need to access to the amplitudes and the relative phases between left and right polarization on the same site to measure the Schmidt norm. The amplitudes of polarization on each site can be recovered by doing the following procedure. After the walk, the resulting state is projected on the $\vert L \rangle$ polarization. Following Ref.~\cite{Cardano2016}, the population of a specific OAM site $\vert j\rangle$ can be read by adding a spatial light modulator (SLM) and a single mode fiber (SMF): The light passes through a hologram generated by the SLM  that  shifts all OAM $\vert k\rangle$ to $\vert k-j\rangle$. Then, the light passes through the SMF that selects only the $\vert 0\rangle$ OAM mode. This allows one to measure the amplitude of the $L$ polarization on each site.The amplitude of the $R$ polarization can be read with a rotation before the projection. The phases $e^{i\phi_j}=\alpha_{jL}^*\alpha_{jR}$ can retrieved by adding different rotation before the projector. The angles $\phi_j$ will be encoded to the amplitude distribution and, then, by measuring the amplitude distribution for each site with the same aforementioned procedure the value of $\phi$ could be extracted.

\section{Conclusions}

We studied the generation of hybrid entanglement in a one-dimensional quantum walk. In particular, we studied the generation of maximally entangled states in a quantum walk with different coin operators. We addressed this problem as  an optimization problem of the Schmidt norm. This algorithm allows one to generate highly entangled states for walks with $10$ steps, with better performances than the Hadamard quantum walks and in a regime where the disordered quantum walks do not converge yet to a maximally entangled state~\cite{Rigolin2013}. We benchmarked the performance of the algorithm for $N=10$ steps, which confirms that the aforementioned states can be prepared with our algorithm for an intermediate number of steps. For $N\gg 10$, the algorithm reaches maximally entangle states but extended states in position are more difficult to reach. However, by adding the inverse participation ratio constraint, the algorithm is still able to reach extended states.

We discussed an experimental implementation in a quantum walk in the OAM and polarization degrees of freedom of light along with the measurement of the Schmidt norm. The number of optical elements scales linearly with the system's size and, currently, state-of-the-art experiments have been performed with around $10$ steps~\cite{Cardano2015,Massignan2017}. It would be interesting to investigate the generation of hybrid entanglement in experimental platforms with other degrees of freedom~\cite{Wang:18}. Furthermore, this problem can be readily generalized to the generation of entanglement in a 2D quantum walk in the setup proposed in Ref.~\cite{Derrico2018}. It would also be interesting to generate highly entangled states with a reinforcement learning algorithm. Finally, with the use of ancillary photons, teleportation of high dimensional states have been accomplished~\cite{Pan2019} and both degrees of freedom of a single photon encoded in the polarization and OAM have been teleported at the same time~\cite{Wang2015}. Therefore, it would be interesting to study the possibility of teleporting such states.

\textit{Acknowledgements.} We thank F. Cardano, R. Chhajlany, M. Maffei and A. Seri for insightful discussions. This work has been supported by the Spanish Ministry MINECO (National Plan 15 Grant: FISICATEAMO No. FIS2016-79508-P, SEVERO OCHOA No. SEV-2015-0522, FPI), European Social Fund, Fundaci\'{o} Cellex, Generalitat de Catalunya (AGAUR Grant No. 2017 SGR 1341 and CERCA/Program), ERC AdG OSYRIS and NOQIA, and the National Science Centre, Poland Symfonia Grant No. 2016/20/W/ST4/00314. A.D. is financed by a Juan de la Cierva fellowship (IJCI-2017-33180).


\begin{thebibliography}{49}%
\makeatletter
\providecommand \@ifxundefined [1]{%
 \@ifx{#1\undefined}
}%
\providecommand \@ifnum [1]{%
 \ifnum #1\expandafter \@firstoftwo
 \else \expandafter \@secondoftwo
 \fi
}%
\providecommand \@ifx [1]{%
 \ifx #1\expandafter \@firstoftwo
 \else \expandafter \@secondoftwo
 \fi
}%
\providecommand \natexlab [1]{#1}%
\providecommand \enquote  [1]{``#1''}%
\providecommand \bibnamefont  [1]{#1}%
\providecommand \bibfnamefont [1]{#1}%
\providecommand \citenamefont [1]{#1}%
\providecommand \href@noop [0]{\@secondoftwo}%
\providecommand \href [0]{\begingroup \@sanitize@url \@href}%
\providecommand \@href[1]{\@@startlink{#1}\@@href}%
\providecommand \@@href[1]{\endgroup#1\@@endlink}%
\providecommand \@sanitize@url [0]{\catcode `\\12\catcode `\$12\catcode
  `\&12\catcode `\#12\catcode `\^12\catcode `\_12\catcode `\%12\relax}%
\providecommand \@@startlink[1]{}%
\providecommand \@@endlink[0]{}%
\providecommand \url  [0]{\begingroup\@sanitize@url \@url }%
\providecommand \@url [1]{\endgroup\@href {#1}{\urlprefix }}%
\providecommand \urlprefix  [0]{URL }%
\providecommand \Eprint [0]{\href }%
\providecommand \doibase [0]{http://dx.doi.org/}%
\providecommand \selectlanguage [0]{\@gobble}%
\providecommand \bibinfo  [0]{\@secondoftwo}%
\providecommand \bibfield  [0]{\@secondoftwo}%
\providecommand \translation [1]{[#1]}%
\providecommand \BibitemOpen [0]{}%
\providecommand \bibitemStop [0]{}%
\providecommand \bibitemNoStop [0]{.\EOS\space}%
\providecommand \EOS [0]{\spacefactor3000\relax}%
\providecommand \BibitemShut  [1]{\csname bibitem#1\endcsname}%
\let\auto@bib@innerbib\@empty
\bibitem [{\citenamefont {Horodecki}\ \emph {et~al.}(2009)\citenamefont
  {Horodecki}, \citenamefont {Horodecki}, \citenamefont {Horodecki},\ and\
  \citenamefont {Horodecki}}]{Horodecki2009}%
  \BibitemOpen
  \bibfield  {author} {\bibinfo {author} {\bibfnamefont {R.}~\bibnamefont
  {Horodecki}}, \bibinfo {author} {\bibfnamefont {P.}~\bibnamefont
  {Horodecki}}, \bibinfo {author} {\bibfnamefont {M.}~\bibnamefont
  {Horodecki}}, \ and\ \bibinfo {author} {\bibfnamefont {K.}~\bibnamefont
  {Horodecki}},\ }\href {\doibase 10.1103/RevModPhys.81.865} {\bibfield
  {journal} {\bibinfo  {journal} {Rev. Mod. Phys.}\ }\textbf {\bibinfo {volume}
  {81}},\ \bibinfo {pages} {865} (\bibinfo {year} {2009})}\BibitemShut
  {NoStop}%
\bibitem [{\citenamefont {Pirandola}\ \emph {et~al.}(2015)\citenamefont
  {Pirandola}, \citenamefont {Eisert}, \citenamefont {Weedbrook}, \citenamefont
  {Furusawa},\ and\ \citenamefont {Braunstein}}]{Braustein2015}%
  \BibitemOpen
  \bibfield  {author} {\bibinfo {author} {\bibfnamefont {S.}~\bibnamefont
  {Pirandola}}, \bibinfo {author} {\bibfnamefont {J.}~\bibnamefont {Eisert}},
  \bibinfo {author} {\bibfnamefont {C.}~\bibnamefont {Weedbrook}}, \bibinfo
  {author} {\bibfnamefont {A.}~\bibnamefont {Furusawa}}, \ and\ \bibinfo
  {author} {\bibfnamefont {S.}~\bibnamefont {Braunstein}},\ }\href
  {https://www.nature.com/articles/nphoton.2015.154} {\bibfield  {journal}
  {\bibinfo  {journal} {Nat. Photonics}\ }\textbf {\bibinfo {volume} {9}},\
  \bibinfo {pages} {641} (\bibinfo {year} {2015})}\BibitemShut {NoStop}%
\bibitem [{\citenamefont {Scarani}\ \emph {et~al.}(2009)\citenamefont
  {Scarani}, \citenamefont {Bechmann-Pasquinucci}, \citenamefont {Cerf},
  \citenamefont {Du\ifmmode~\check{s}\else \v{s}\fi{}ek}, \citenamefont
  {L\"utkenhaus},\ and\ \citenamefont {Peev}}]{Scarani2009}%
  \BibitemOpen
  \bibfield  {author} {\bibinfo {author} {\bibfnamefont {V.}~\bibnamefont
  {Scarani}}, \bibinfo {author} {\bibfnamefont {H.}~\bibnamefont
  {Bechmann-Pasquinucci}}, \bibinfo {author} {\bibfnamefont {N.~J.}\
  \bibnamefont {Cerf}}, \bibinfo {author} {\bibfnamefont {M.}~\bibnamefont
  {Du\ifmmode~\check{s}\else \v{s}\fi{}ek}}, \bibinfo {author} {\bibfnamefont
  {N.}~\bibnamefont {L\"utkenhaus}}, \ and\ \bibinfo {author} {\bibfnamefont
  {M.}~\bibnamefont {Peev}},\ }\href {\doibase 10.1103/RevModPhys.81.1301}
  {\bibfield  {journal} {\bibinfo  {journal} {Rev. Mod. Phys.}\ }\textbf
  {\bibinfo {volume} {81}},\ \bibinfo {pages} {1301} (\bibinfo {year}
  {2009})}\BibitemShut {NoStop}%
\bibitem [{\citenamefont {Rosset}\ \emph {et~al.}(2014)\citenamefont {Rosset},
  \citenamefont {Bancal},\ and\ \citenamefont {Gisin}}]{Rosset_2014}%
  \BibitemOpen
  \bibfield  {author} {\bibinfo {author} {\bibfnamefont {D.}~\bibnamefont
  {Rosset}}, \bibinfo {author} {\bibfnamefont {J.-D.}\ \bibnamefont {Bancal}},
  \ and\ \bibinfo {author} {\bibfnamefont {N.}~\bibnamefont {Gisin}},\ }\href
  {\doibase 10.1088/1751-8113/47/42/424022} {\bibfield  {journal} {\bibinfo
  {journal} {J Phys A-Math Theor}\ }\textbf {\bibinfo {volume} {47}},\ \bibinfo
  {pages} {424022} (\bibinfo {year} {2014})}\BibitemShut {NoStop}%
\bibitem [{\citenamefont {Brunner}\ \emph
  {et~al.}(2014{\natexlab{a}})\citenamefont {Brunner}, \citenamefont
  {G{\"u}hne},\ and\ \citenamefont {Huber}}]{Brunner_2014}%
  \BibitemOpen
  \bibfield  {author} {\bibinfo {author} {\bibfnamefont {N.}~\bibnamefont
  {Brunner}}, \bibinfo {author} {\bibfnamefont {O.}~\bibnamefont {G{\"u}hne}},
  \ and\ \bibinfo {author} {\bibfnamefont {M.}~\bibnamefont {Huber}},\ }\href
  {\doibase 10.1088/1751-8113/47/42/420301} {\bibfield  {journal} {\bibinfo
  {journal} {J Phys A-Math Theor}\ }\textbf {\bibinfo {volume} {47}},\ \bibinfo
  {pages} {420301} (\bibinfo {year} {2014}{\natexlab{a}})}\BibitemShut
  {NoStop}%
\bibitem [{\citenamefont {Brunner}\ \emph
  {et~al.}(2014{\natexlab{b}})\citenamefont {Brunner}, \citenamefont
  {Cavalcanti}, \citenamefont {Pironio}, \citenamefont {Scarani},\ and\
  \citenamefont {Wehner}}]{Brunner2014}%
  \BibitemOpen
  \bibfield  {author} {\bibinfo {author} {\bibfnamefont {N.}~\bibnamefont
  {Brunner}}, \bibinfo {author} {\bibfnamefont {D.}~\bibnamefont {Cavalcanti}},
  \bibinfo {author} {\bibfnamefont {S.}~\bibnamefont {Pironio}}, \bibinfo
  {author} {\bibfnamefont {V.}~\bibnamefont {Scarani}}, \ and\ \bibinfo
  {author} {\bibfnamefont {S.}~\bibnamefont {Wehner}},\ }\href {\doibase
  10.1103/RevModPhys.86.419} {\bibfield  {journal} {\bibinfo  {journal} {Rev.
  Mod. Phys.}\ }\textbf {\bibinfo {volume} {86}},\ \bibinfo {pages} {419}
  (\bibinfo {year} {2014}{\natexlab{b}})}\BibitemShut {NoStop}%
\bibitem [{\citenamefont {Krenn}\ \emph {et~al.}(2016)\citenamefont {Krenn},
  \citenamefont {Malik}, \citenamefont {Erhard},\ and\ \citenamefont
  {Zeilinger}}]{Zeilinger2017}%
  \BibitemOpen
  \bibfield  {author} {\bibinfo {author} {\bibfnamefont {M.}~\bibnamefont
  {Krenn}}, \bibinfo {author} {\bibfnamefont {M.}~\bibnamefont {Malik}},
  \bibinfo {author} {\bibfnamefont {M.}~\bibnamefont {Erhard}}, \ and\ \bibinfo
  {author} {\bibfnamefont {A.}~\bibnamefont {Zeilinger}},\ }\href
  {https://royalsocietypublishing.org/doi/10.1098/rsta.2015.0442} {\bibfield
  {journal} {\bibinfo  {journal} {Philos. Trans. Royal Soc. A}\ }\textbf
  {\bibinfo {volume} {375}} (\bibinfo {year} {2016})}\BibitemShut {NoStop}%
\bibitem [{\citenamefont {Can}\ \emph {et~al.}(2005)\citenamefont {Can},
  \citenamefont {Klyachko},\ and\ \citenamefont {Shumovsky}}]{Can2005}%
  \BibitemOpen
  \bibfield  {author} {\bibinfo {author} {\bibfnamefont {M.~A.}\ \bibnamefont
  {Can}}, \bibinfo {author} {\bibfnamefont {A.}~\bibnamefont {Klyachko}}, \
  and\ \bibinfo {author} {\bibfnamefont {A.}~\bibnamefont {Shumovsky}},\ }\href
  {http://stacks.iop.org/1464-4266/7/i=2/a=L01} {\bibfield  {journal} {\bibinfo
   {journal} {J. Opt B Quantum Semiclassical Opt.}\ }\textbf {\bibinfo {volume}
  {7}},\ \bibinfo {pages} {L1} (\bibinfo {year} {2005})}\BibitemShut {NoStop}%
\bibitem [{\citenamefont {Aiello}\ \emph {et~al.}(2015)\citenamefont {Aiello},
  \citenamefont {Tšppel}, \citenamefont {Marquardt}, \citenamefont
  {Giacobino},\ and\ \citenamefont {Leuchs}}]{Aiello_2015}%
  \BibitemOpen
  \bibfield  {author} {\bibinfo {author} {\bibfnamefont {A.}~\bibnamefont
  {Aiello}}, \bibinfo {author} {\bibfnamefont {F.}~\bibnamefont {Tšppel}},
  \bibinfo {author} {\bibfnamefont {C.}~\bibnamefont {Marquardt}}, \bibinfo
  {author} {\bibfnamefont {E.}~\bibnamefont {Giacobino}}, \ and\ \bibinfo
  {author} {\bibfnamefont {G.}~\bibnamefont {Leuchs}},\ }\href {\doibase
  10.1088/1367-2630/17/4/043024} {\bibfield  {journal} {\bibinfo  {journal}
  {New Journal of Physics}\ }\textbf {\bibinfo {volume} {17}},\ \bibinfo
  {pages} {043024} (\bibinfo {year} {2015})}\BibitemShut {NoStop}%
\bibitem [{\citenamefont {Karimi}\ and\ \citenamefont
  {Boyd}(2015)}]{Karimi1172}%
  \BibitemOpen
  \bibfield  {author} {\bibinfo {author} {\bibfnamefont {E.}~\bibnamefont
  {Karimi}}\ and\ \bibinfo {author} {\bibfnamefont {R.~W.}\ \bibnamefont
  {Boyd}},\ }\href {\doibase 10.1126/science.aad7174} {\bibfield  {journal}
  {\bibinfo  {journal} {Science}\ }\textbf {\bibinfo {volume} {350}},\ \bibinfo
  {pages} {1172} (\bibinfo {year} {2015})}\BibitemShut {NoStop}%
\bibitem [{\citenamefont {Karimi}\ \emph {et~al.}(2010)\citenamefont {Karimi},
  \citenamefont {Leach}, \citenamefont {Slussarenko}, \citenamefont
  {Piccirillo}, \citenamefont {Marrucci}, \citenamefont {Chen}, \citenamefont
  {She}, \citenamefont {Franke-Arnold}, \citenamefont {Padgett},\ and\
  \citenamefont {Santamato}}]{Karimi}%
  \BibitemOpen
  \bibfield  {author} {\bibinfo {author} {\bibfnamefont {E.}~\bibnamefont
  {Karimi}}, \bibinfo {author} {\bibfnamefont {J.}~\bibnamefont {Leach}},
  \bibinfo {author} {\bibfnamefont {S.}~\bibnamefont {Slussarenko}}, \bibinfo
  {author} {\bibfnamefont {B.}~\bibnamefont {Piccirillo}}, \bibinfo {author}
  {\bibfnamefont {L.}~\bibnamefont {Marrucci}}, \bibinfo {author}
  {\bibfnamefont {L.}~\bibnamefont {Chen}}, \bibinfo {author} {\bibfnamefont
  {W.}~\bibnamefont {She}}, \bibinfo {author} {\bibfnamefont {S.}~\bibnamefont
  {Franke-Arnold}}, \bibinfo {author} {\bibfnamefont {M.~J.}\ \bibnamefont
  {Padgett}}, \ and\ \bibinfo {author} {\bibfnamefont {E.}~\bibnamefont
  {Santamato}},\ }\href {\doibase 10.1103/PhysRevA.82.022115} {\bibfield
  {journal} {\bibinfo  {journal} {Phys. Rev. A}\ }\textbf {\bibinfo {volume}
  {82}},\ \bibinfo {pages} {022115} (\bibinfo {year} {2010})}\BibitemShut
  {NoStop}%
\bibitem [{\citenamefont {Flamini}\ \emph {et~al.}(2018)\citenamefont
  {Flamini}, \citenamefont {Spagnolo},\ and\ \citenamefont
  {Sciarrino}}]{Sciarrino2018}%
  \BibitemOpen
  \bibfield  {author} {\bibinfo {author} {\bibfnamefont {F.}~\bibnamefont
  {Flamini}}, \bibinfo {author} {\bibfnamefont {N.}~\bibnamefont {Spagnolo}}, \
  and\ \bibinfo {author} {\bibfnamefont {F.}~\bibnamefont {Sciarrino}},\ }\href
  {\doibase 10.1088/1361-6633/aad5b2} {\bibfield  {journal} {\bibinfo
  {journal} {Rep. Prog. Phys.}\ }\textbf {\bibinfo {volume} {82}},\ \bibinfo
  {pages} {016001} (\bibinfo {year} {2018})}\BibitemShut {NoStop}%
\bibitem [{\citenamefont {Rubinsztein-Dunlop}\ and\ \citenamefont
  {et~al.}(2017)}]{Karimi2017}%
  \BibitemOpen
  \bibfield  {author} {\bibinfo {author} {\bibfnamefont {H.}~\bibnamefont
  {Rubinsztein-Dunlop}}\ and\ \bibinfo {author} {\bibnamefont {et~al.}},\
  }\href@noop {} {\bibfield  {journal} {\bibinfo  {journal} {Journal of
  Optics}\ }\textbf {\bibinfo {volume} {19}} (\bibinfo {year}
  {2017})}\BibitemShut {NoStop}%
\bibitem [{\citenamefont {L\"u}\ \emph {et~al.}(2018)\citenamefont {L\"u},
  \citenamefont {Zhu}, \citenamefont {Zheng},\ and\ \citenamefont
  {Wu}}]{Ying2018}%
  \BibitemOpen
  \bibfield  {author} {\bibinfo {author} {\bibfnamefont {X.-Y.}\ \bibnamefont
  {L\"u}}, \bibinfo {author} {\bibfnamefont {G.-L.}\ \bibnamefont {Zhu}},
  \bibinfo {author} {\bibfnamefont {L.-L.}\ \bibnamefont {Zheng}}, \ and\
  \bibinfo {author} {\bibfnamefont {Y.}~\bibnamefont {Wu}},\ }\href {\doibase
  10.1103/PhysRevA.97.033807} {\bibfield  {journal} {\bibinfo  {journal} {Phys.
  Rev. A}\ }\textbf {\bibinfo {volume} {97}},\ \bibinfo {pages} {033807}
  (\bibinfo {year} {2018})}\BibitemShut {NoStop}%
\bibitem [{\citenamefont {Zhang}\ \emph {et~al.}(2018)\citenamefont {Zhang},
  \citenamefont {Zhang}, \citenamefont {Cen}, \citenamefont {You},
  \citenamefont {Adhikari}, \citenamefont {Dowling},\ and\ \citenamefont
  {Zhao}}]{Zhang:18}%
  \BibitemOpen
  \bibfield  {author} {\bibinfo {author} {\bibfnamefont {J.-D.}\ \bibnamefont
  {Zhang}}, \bibinfo {author} {\bibfnamefont {Z.-J.}\ \bibnamefont {Zhang}},
  \bibinfo {author} {\bibfnamefont {L.-Z.}\ \bibnamefont {Cen}}, \bibinfo
  {author} {\bibfnamefont {C.}~\bibnamefont {You}}, \bibinfo {author}
  {\bibfnamefont {S.}~\bibnamefont {Adhikari}}, \bibinfo {author}
  {\bibfnamefont {J.~P.}\ \bibnamefont {Dowling}}, \ and\ \bibinfo {author}
  {\bibfnamefont {Y.}~\bibnamefont {Zhao}},\ }\href {\doibase
  10.1364/OE.26.016524} {\bibfield  {journal} {\bibinfo  {journal} {Opt.
  Express}\ }\textbf {\bibinfo {volume} {26}},\ \bibinfo {pages} {16524}
  (\bibinfo {year} {2018})}\BibitemShut {NoStop}%
\bibitem [{\citenamefont {Wang}\ \emph {et~al.}(2015)\citenamefont {Wang},
  \citenamefont {Cai}, \citenamefont {Su}, \citenamefont {Chen}, \citenamefont
  {Wu}, \citenamefont {Li}, \citenamefont {Liu}, \citenamefont {Lu},\ and\
  \citenamefont {Pan}}]{Wang2015}%
  \BibitemOpen
  \bibfield  {author} {\bibinfo {author} {\bibfnamefont {X.-L.}\ \bibnamefont
  {Wang}}, \bibinfo {author} {\bibfnamefont {X.-D.}\ \bibnamefont {Cai}},
  \bibinfo {author} {\bibfnamefont {Z.-E.}\ \bibnamefont {Su}}, \bibinfo
  {author} {\bibfnamefont {M.-c.}\ \bibnamefont {Chen}}, \bibinfo {author}
  {\bibfnamefont {D.}~\bibnamefont {Wu}}, \bibinfo {author} {\bibfnamefont
  {L.}~\bibnamefont {Li}}, \bibinfo {author} {\bibfnamefont {N.-L.}\
  \bibnamefont {Liu}}, \bibinfo {author} {\bibfnamefont {C.-Y.}\ \bibnamefont
  {Lu}}, \ and\ \bibinfo {author} {\bibfnamefont {J.-W.}\ \bibnamefont {Pan}},\
  }\href {\doibase 10.1038/nature14246} {\bibfield  {journal} {\bibinfo
  {journal} {Nature}\ }\textbf {\bibinfo {volume} {518}},\ \bibinfo {pages}
  {516} (\bibinfo {year} {2015})}\BibitemShut {NoStop}%
\bibitem [{\citenamefont {Vitelli}\ \emph {et~al.}(2013)\citenamefont
  {Vitelli}, \citenamefont {Spagnolo}, \citenamefont {Aparo}, \citenamefont
  {Sciarrino}, \citenamefont {Santamato},\ and\ \citenamefont
  {Marrucci}}]{Vitelli2013}%
  \BibitemOpen
  \bibfield  {author} {\bibinfo {author} {\bibfnamefont {C.}~\bibnamefont
  {Vitelli}}, \bibinfo {author} {\bibfnamefont {N.}~\bibnamefont {Spagnolo}},
  \bibinfo {author} {\bibfnamefont {L.}~\bibnamefont {Aparo}}, \bibinfo
  {author} {\bibfnamefont {F.}~\bibnamefont {Sciarrino}}, \bibinfo {author}
  {\bibfnamefont {E.}~\bibnamefont {Santamato}}, \ and\ \bibinfo {author}
  {\bibfnamefont {L.}~\bibnamefont {Marrucci}},\ }\href
  {https://doi.org/10.1038/nphoton.2013.107} {\bibfield  {journal} {\bibinfo
  {journal} {Nat. Photonics}\ }\textbf {\bibinfo {volume} {7}},\ \bibinfo
  {pages} {521} (\bibinfo {year} {2013})}\BibitemShut {NoStop}%
\bibitem [{\citenamefont {Erhard}\ \emph {et~al.}(2017)\citenamefont {Erhard},
  \citenamefont {Fickler}, \citenamefont {Krenn},\ and\ \citenamefont
  {Zeilinger}}]{Zeilinger2018}%
  \BibitemOpen
  \bibfield  {author} {\bibinfo {author} {\bibfnamefont {M.}~\bibnamefont
  {Erhard}}, \bibinfo {author} {\bibfnamefont {R.}~\bibnamefont {Fickler}},
  \bibinfo {author} {\bibfnamefont {M.}~\bibnamefont {Krenn}}, \ and\ \bibinfo
  {author} {\bibfnamefont {A.}~\bibnamefont {Zeilinger}},\ }\href
  {https://www.nature.com/articles/lsa2017146} {\bibfield  {journal} {\bibinfo
  {journal} {Light Sci Appl.}\ }\textbf {\bibinfo {volume} {7}},\ \bibinfo
  {pages} {17146} (\bibinfo {year} {2017})}\BibitemShut {NoStop}%
\bibitem [{\citenamefont {Deng}\ \emph {et~al.}(2007)\citenamefont {Deng},
  \citenamefont {Wang},\ and\ \citenamefont {Wang}}]{Deng2007}%
  \BibitemOpen
  \bibfield  {author} {\bibinfo {author} {\bibfnamefont {L.-P.}\ \bibnamefont
  {Deng}}, \bibinfo {author} {\bibfnamefont {H.}~\bibnamefont {Wang}}, \ and\
  \bibinfo {author} {\bibfnamefont {K.}~\bibnamefont {Wang}},\ }\href {\doibase
  10.1364/JOSAB.24.002517} {\bibfield  {journal} {\bibinfo  {journal} {J. Opt.
  Soc. Am. B}\ }\textbf {\bibinfo {volume} {24}},\ \bibinfo {pages} {2517}
  (\bibinfo {year} {2007})}\BibitemShut {NoStop}%
\bibitem [{\citenamefont {Lopes}\ \emph {et~al.}(2018)\citenamefont {Lopes},
  \citenamefont {Soares}, \citenamefont {Bernardo}, \citenamefont {Caetano},\
  and\ \citenamefont {Canabarro}}]{Canabarro2018}%
  \BibitemOpen
  \bibfield  {author} {\bibinfo {author} {\bibfnamefont {J.}~\bibnamefont
  {Lopes}}, \bibinfo {author} {\bibfnamefont {W.}~\bibnamefont {Soares}},
  \bibinfo {author} {\bibfnamefont {B.}~\bibnamefont {Bernardo}}, \bibinfo
  {author} {\bibfnamefont {D.}~\bibnamefont {Caetano}}, \ and\ \bibinfo
  {author} {\bibfnamefont {A.}~\bibnamefont {Canabarro}},\ }\href@noop {} {\ ,\
  \bibinfo {eid} {arXiv:1811.04001} (\bibinfo {year} {2018})},\ \Eprint
  {http://arxiv.org/abs/1811.04001} {1811.04001} \BibitemShut {NoStop}%
\bibitem [{\citenamefont {Nielsen}\ and\ \citenamefont {Chuang}(2011)}]{NC}%
  \BibitemOpen
  \bibfield  {author} {\bibinfo {author} {\bibfnamefont {M.~A.}\ \bibnamefont
  {Nielsen}}\ and\ \bibinfo {author} {\bibfnamefont {I.~L.}\ \bibnamefont
  {Chuang}},\ }\href@noop {} {\emph {\bibinfo {title} {Quantum Computation and
  Quantum Information: 10th Anniversary Edition}}},\ \bibinfo {edition} {10th}\
  ed.\ (\bibinfo  {publisher} {Cambridge University Press},\ \bibinfo {address}
  {New York, NY, USA},\ \bibinfo {year} {2011})\BibitemShut {NoStop}%
\bibitem [{\citenamefont {Cozzolino}\ \emph
  {et~al.}(2019{\natexlab{a}})\citenamefont {Cozzolino}, \citenamefont {Bacco},
  \citenamefont {Da~Lio}, \citenamefont {Ingerslev}, \citenamefont {Ding},
  \citenamefont {Dalgaard}, \citenamefont {Kristensen}, \citenamefont {Galili},
  \citenamefont {Rottwitt}, \citenamefont {Ramachandran},\ and\ \citenamefont
  {Oxenl\o{}we}}]{Cozzolino2018}%
  \BibitemOpen
  \bibfield  {author} {\bibinfo {author} {\bibfnamefont {D.}~\bibnamefont
  {Cozzolino}}, \bibinfo {author} {\bibfnamefont {D.}~\bibnamefont {Bacco}},
  \bibinfo {author} {\bibfnamefont {B.}~\bibnamefont {Da~Lio}}, \bibinfo
  {author} {\bibfnamefont {K.}~\bibnamefont {Ingerslev}}, \bibinfo {author}
  {\bibfnamefont {Y.}~\bibnamefont {Ding}}, \bibinfo {author} {\bibfnamefont
  {K.}~\bibnamefont {Dalgaard}}, \bibinfo {author} {\bibfnamefont
  {P.}~\bibnamefont {Kristensen}}, \bibinfo {author} {\bibfnamefont
  {M.}~\bibnamefont {Galili}}, \bibinfo {author} {\bibfnamefont
  {K.}~\bibnamefont {Rottwitt}}, \bibinfo {author} {\bibfnamefont
  {S.}~\bibnamefont {Ramachandran}}, \ and\ \bibinfo {author} {\bibfnamefont
  {L.~K.}\ \bibnamefont {Oxenl\o{}we}},\ }\href {\doibase
  10.1103/PhysRevApplied.11.064058} {\bibfield  {journal} {\bibinfo  {journal}
  {Phys. Rev. Applied}\ }\textbf {\bibinfo {volume} {11}},\ \bibinfo {pages}
  {064058} (\bibinfo {year} {2019}{\natexlab{a}})}\BibitemShut {NoStop}%
\bibitem [{\citenamefont {Cozzolino}\ \emph
  {et~al.}(2019{\natexlab{b}})\citenamefont {Cozzolino}, \citenamefont
  {Polino}, \citenamefont {Valeri}, \citenamefont {Carvacho}, \citenamefont
  {Bacco}, \citenamefont {Spagnolo}, \citenamefont {Oxenl¿we},\ and\
  \citenamefont {Sciarrino}}]{Cozzolino2019}%
  \BibitemOpen
  \bibfield  {author} {\bibinfo {author} {\bibfnamefont {D.}~\bibnamefont
  {Cozzolino}}, \bibinfo {author} {\bibfnamefont {E.}~\bibnamefont {Polino}},
  \bibinfo {author} {\bibfnamefont {M.}~\bibnamefont {Valeri}}, \bibinfo
  {author} {\bibfnamefont {G.}~\bibnamefont {Carvacho}}, \bibinfo {author}
  {\bibfnamefont {D.}~\bibnamefont {Bacco}}, \bibinfo {author} {\bibfnamefont
  {N.}~\bibnamefont {Spagnolo}}, \bibinfo {author} {\bibfnamefont {L.~K.~K.}\
  \bibnamefont {Oxenl¿we}}, \ and\ \bibinfo {author} {\bibfnamefont
  {F.}~\bibnamefont {Sciarrino}},\ }\href {\doibase 10.1117/1.AP.1.4.046005}
  {\bibfield  {journal} {\bibinfo  {journal} {Adv. Photon.}\ }\textbf {\bibinfo
  {volume} {1}},\ \bibinfo {pages} {1 } (\bibinfo {year}
  {2019}{\natexlab{b}})}\BibitemShut {NoStop}%
\bibitem [{\citenamefont {Manouchehri}\ and\ \citenamefont
  {Wang}(2013)}]{Wang_book}%
  \BibitemOpen
  \bibfield  {author} {\bibinfo {author} {\bibfnamefont {K.}~\bibnamefont
  {Manouchehri}}\ and\ \bibinfo {author} {\bibfnamefont {J.}~\bibnamefont
  {Wang}},\ }\href@noop {} {\emph {\bibinfo {title} {Physical Implementation of
  Quantum Walks}}}\ (\bibinfo  {publisher} {Springer Publishing Company,
  Incorporated},\ \bibinfo {year} {2013})\BibitemShut {NoStop}%
\bibitem [{\citenamefont {Karski}\ \emph {et~al.}(2009)\citenamefont {Karski},
  \citenamefont {F{\"o}rster}, \citenamefont {Choi}, \citenamefont {Steffen},
  \citenamefont {Alt}, \citenamefont {Meschede},\ and\ \citenamefont
  {Widera}}]{Widera2009}%
  \BibitemOpen
  \bibfield  {author} {\bibinfo {author} {\bibfnamefont {M.}~\bibnamefont
  {Karski}}, \bibinfo {author} {\bibfnamefont {L.}~\bibnamefont {F{\"o}rster}},
  \bibinfo {author} {\bibfnamefont {J.-M.}\ \bibnamefont {Choi}}, \bibinfo
  {author} {\bibfnamefont {A.}~\bibnamefont {Steffen}}, \bibinfo {author}
  {\bibfnamefont {W.}~\bibnamefont {Alt}}, \bibinfo {author} {\bibfnamefont
  {D.}~\bibnamefont {Meschede}}, \ and\ \bibinfo {author} {\bibfnamefont
  {A.}~\bibnamefont {Widera}},\ }\href {\doibase 10.1126/science.1174436}
  {\bibfield  {journal} {\bibinfo  {journal} {Science}\ }\textbf {\bibinfo
  {volume} {325}},\ \bibinfo {pages} {174} (\bibinfo {year}
  {2009})}\BibitemShut {NoStop}%
\bibitem [{\citenamefont {Z\"ahringer}\ \emph {et~al.}(2010)\citenamefont
  {Z\"ahringer}, \citenamefont {Kirchmair}, \citenamefont {Gerritsma},
  \citenamefont {Solano}, \citenamefont {Blatt},\ and\ \citenamefont
  {Roos}}]{Roos}%
  \BibitemOpen
  \bibfield  {author} {\bibinfo {author} {\bibfnamefont {F.}~\bibnamefont
  {Z\"ahringer}}, \bibinfo {author} {\bibfnamefont {G.}~\bibnamefont
  {Kirchmair}}, \bibinfo {author} {\bibfnamefont {R.}~\bibnamefont
  {Gerritsma}}, \bibinfo {author} {\bibfnamefont {E.}~\bibnamefont {Solano}},
  \bibinfo {author} {\bibfnamefont {R.}~\bibnamefont {Blatt}}, \ and\ \bibinfo
  {author} {\bibfnamefont {C.~F.}\ \bibnamefont {Roos}},\ }\href {\doibase
  10.1103/PhysRevLett.104.100503} {\bibfield  {journal} {\bibinfo  {journal}
  {Phys. Rev. Lett.}\ }\textbf {\bibinfo {volume} {104}},\ \bibinfo {pages}
  {100503} (\bibinfo {year} {2010})}\BibitemShut {NoStop}%
\bibitem [{\citenamefont {Schmitz}\ \emph {et~al.}(2009)\citenamefont
  {Schmitz}, \citenamefont {Matjeschk}, \citenamefont {Schneider},
  \citenamefont {Glueckert}, \citenamefont {Enderlein}, \citenamefont {Huber},\
  and\ \citenamefont {Schaetz}}]{Schmitz}%
  \BibitemOpen
  \bibfield  {author} {\bibinfo {author} {\bibfnamefont {H.}~\bibnamefont
  {Schmitz}}, \bibinfo {author} {\bibfnamefont {R.}~\bibnamefont {Matjeschk}},
  \bibinfo {author} {\bibfnamefont {C.}~\bibnamefont {Schneider}}, \bibinfo
  {author} {\bibfnamefont {J.}~\bibnamefont {Glueckert}}, \bibinfo {author}
  {\bibfnamefont {M.}~\bibnamefont {Enderlein}}, \bibinfo {author}
  {\bibfnamefont {T.}~\bibnamefont {Huber}}, \ and\ \bibinfo {author}
  {\bibfnamefont {T.}~\bibnamefont {Schaetz}},\ }\href {\doibase
  10.1103/PhysRevLett.103.090504} {\bibfield  {journal} {\bibinfo  {journal}
  {Phys. Rev. Lett.}\ }\textbf {\bibinfo {volume} {103}},\ \bibinfo {pages}
  {090504} (\bibinfo {year} {2009})}\BibitemShut {NoStop}%
\bibitem [{\citenamefont {Meinert}\ \emph {et~al.}(2014)\citenamefont
  {Meinert}, \citenamefont {Mark}, \citenamefont {Kirilov}, \citenamefont
  {Lauber}, \citenamefont {Weinmann}, \citenamefont {Gr{\"o}bner},
  \citenamefont {Daley},\ and\ \citenamefont {N{\"a}gerl}}]{Meinert}%
  \BibitemOpen
  \bibfield  {author} {\bibinfo {author} {\bibfnamefont {F.}~\bibnamefont
  {Meinert}}, \bibinfo {author} {\bibfnamefont {M.~J.}\ \bibnamefont {Mark}},
  \bibinfo {author} {\bibfnamefont {E.}~\bibnamefont {Kirilov}}, \bibinfo
  {author} {\bibfnamefont {K.}~\bibnamefont {Lauber}}, \bibinfo {author}
  {\bibfnamefont {P.}~\bibnamefont {Weinmann}}, \bibinfo {author}
  {\bibfnamefont {M.}~\bibnamefont {Gr{\"o}bner}}, \bibinfo {author}
  {\bibfnamefont {A.~J.}\ \bibnamefont {Daley}}, \ and\ \bibinfo {author}
  {\bibfnamefont {H.-C.}\ \bibnamefont {N{\"a}gerl}},\ }\href {\doibase
  10.1126/science.1248402} {\bibfield  {journal} {\bibinfo  {journal}
  {Science}\ }\textbf {\bibinfo {volume} {344}},\ \bibinfo {pages} {1259}
  (\bibinfo {year} {2014})}\BibitemShut {NoStop}%
\bibitem [{\citenamefont {Broome}\ \emph {et~al.}(2010)\citenamefont {Broome},
  \citenamefont {Fedrizzi}, \citenamefont {Lanyon}, \citenamefont {Kassal},
  \citenamefont {Aspuru-Guzik},\ and\ \citenamefont {White}}]{Broome2010}%
  \BibitemOpen
  \bibfield  {author} {\bibinfo {author} {\bibfnamefont {M.~A.}\ \bibnamefont
  {Broome}}, \bibinfo {author} {\bibfnamefont {A.}~\bibnamefont {Fedrizzi}},
  \bibinfo {author} {\bibfnamefont {B.~P.}\ \bibnamefont {Lanyon}}, \bibinfo
  {author} {\bibfnamefont {I.}~\bibnamefont {Kassal}}, \bibinfo {author}
  {\bibfnamefont {A.}~\bibnamefont {Aspuru-Guzik}}, \ and\ \bibinfo {author}
  {\bibfnamefont {A.~G.}\ \bibnamefont {White}},\ }\href {\doibase
  10.1103/PhysRevLett.104.153602} {\bibfield  {journal} {\bibinfo  {journal}
  {Phys. Rev. Lett.}\ }\textbf {\bibinfo {volume} {104}},\ \bibinfo {pages}
  {153602} (\bibinfo {year} {2010})}\BibitemShut {NoStop}%
\bibitem [{\citenamefont {Schreiber}\ \emph {et~al.}(2010)\citenamefont
  {Schreiber}, \citenamefont {Cassemiro}, \citenamefont
  {Poto\ifmmode~\check{c}\else \v{c}\fi{}ek}, \citenamefont {G\'abris},
  \citenamefont {Mosley}, \citenamefont {Andersson}, \citenamefont {Jex},\ and\
  \citenamefont {Silberhorn}}]{Schreiber2010}%
  \BibitemOpen
  \bibfield  {author} {\bibinfo {author} {\bibfnamefont {A.}~\bibnamefont
  {Schreiber}}, \bibinfo {author} {\bibfnamefont {K.~N.}\ \bibnamefont
  {Cassemiro}}, \bibinfo {author} {\bibfnamefont {V.}~\bibnamefont
  {Poto\ifmmode~\check{c}\else \v{c}\fi{}ek}}, \bibinfo {author} {\bibfnamefont
  {A.}~\bibnamefont {G\'abris}}, \bibinfo {author} {\bibfnamefont {P.~J.}\
  \bibnamefont {Mosley}}, \bibinfo {author} {\bibfnamefont {E.}~\bibnamefont
  {Andersson}}, \bibinfo {author} {\bibfnamefont {I.}~\bibnamefont {Jex}}, \
  and\ \bibinfo {author} {\bibfnamefont {C.}~\bibnamefont {Silberhorn}},\
  }\href {\doibase 10.1103/PhysRevLett.104.050502} {\bibfield  {journal}
  {\bibinfo  {journal} {Phys. Rev. Lett.}\ }\textbf {\bibinfo {volume} {104}},\
  \bibinfo {pages} {050502} (\bibinfo {year} {2010})}\BibitemShut {NoStop}%
\bibitem [{\citenamefont {Cardano}\ and\ \citenamefont
  {et~al.}(2015)}]{Cardano2015}%
  \BibitemOpen
  \bibfield  {author} {\bibinfo {author} {\bibfnamefont {F.}~\bibnamefont
  {Cardano}}\ and\ \bibinfo {author} {\bibnamefont {et~al.}},\ }\href
  {https://advances.sciencemag.org/content/1/2/e1500087} {\bibfield  {journal}
  {\bibinfo  {journal} {Sci Adv.}\ }\textbf {\bibinfo {volume} {1}},\ \bibinfo
  {pages} {e1500087} (\bibinfo {year} {2015})}\BibitemShut {NoStop}%
\bibitem [{\citenamefont {Barz}\ \emph {et~al.}(2014)\citenamefont {Barz},
  \citenamefont {Kassal}, \citenamefont {Ringbauer}, \citenamefont {Ole~Lipp},
  \citenamefont {Daki{\'c}}, \citenamefont {Aspuru-Guzik},\ and\ \citenamefont
  {Walther}}]{Walther2014}%
  \BibitemOpen
  \bibfield  {author} {\bibinfo {author} {\bibfnamefont {S.}~\bibnamefont
  {Barz}}, \bibinfo {author} {\bibfnamefont {I.}~\bibnamefont {Kassal}},
  \bibinfo {author} {\bibfnamefont {M.}~\bibnamefont {Ringbauer}}, \bibinfo
  {author} {\bibfnamefont {Y.}~\bibnamefont {Ole~Lipp}}, \bibinfo {author}
  {\bibfnamefont {B.}~\bibnamefont {Daki{\'c}}}, \bibinfo {author}
  {\bibfnamefont {A.}~\bibnamefont {Aspuru-Guzik}}, \ and\ \bibinfo {author}
  {\bibfnamefont {P.}~\bibnamefont {Walther}},\ }\href {\doibase
  10.1038/srep06115} {\bibfield  {journal} {\bibinfo  {journal} {Sci Rep.}\
  }\textbf {\bibinfo {volume} {4}},\ \bibinfo {pages} {6115} (\bibinfo {year}
  {2014})}\BibitemShut {NoStop}%
\bibitem [{\citenamefont {Walther}\ \emph {et~al.}(2005)\citenamefont
  {Walther}, \citenamefont {J~Resch}, \citenamefont {Rudolph}, \citenamefont
  {Schenck}, \citenamefont {Weinfurter}, \citenamefont {Vedral}, \citenamefont
  {Aspelmeyer},\ and\ \citenamefont {Zeilinger}}]{Zeilinger2005}%
  \BibitemOpen
  \bibfield  {author} {\bibinfo {author} {\bibfnamefont {P.}~\bibnamefont
  {Walther}}, \bibinfo {author} {\bibfnamefont {K.}~\bibnamefont {J~Resch}},
  \bibinfo {author} {\bibfnamefont {T.}~\bibnamefont {Rudolph}}, \bibinfo
  {author} {\bibfnamefont {E.}~\bibnamefont {Schenck}}, \bibinfo {author}
  {\bibfnamefont {H.}~\bibnamefont {Weinfurter}}, \bibinfo {author}
  {\bibfnamefont {V.}~\bibnamefont {Vedral}}, \bibinfo {author} {\bibfnamefont
  {M.}~\bibnamefont {Aspelmeyer}}, \ and\ \bibinfo {author} {\bibfnamefont
  {A.}~\bibnamefont {Zeilinger}},\ }\href {\doibase 10.1038/nature03347}
  {\bibfield  {journal} {\bibinfo  {journal} {Nature}\ }\textbf {\bibinfo
  {volume} {434}},\ \bibinfo {pages} {169} (\bibinfo {year}
  {2005})}\BibitemShut {NoStop}%
\bibitem [{\citenamefont {Cardano}\ and\ \citenamefont
  {et~al.}(2017)}]{Massignan2017}%
  \BibitemOpen
  \bibfield  {author} {\bibinfo {author} {\bibfnamefont {F.}~\bibnamefont
  {Cardano}}\ and\ \bibinfo {author} {\bibnamefont {et~al.}},\ }\href
  {http://europepmc.org/articles/PMC5501976} {\bibfield  {journal} {\bibinfo
  {journal} {Nat. Commun.}\ }\textbf {\bibinfo {volume} {8}},\ \bibinfo {pages}
  {15516} (\bibinfo {year} {2017})}\BibitemShut {NoStop}%
\bibitem [{\citenamefont {Zhang}\ \emph {et~al.}(2016)\citenamefont {Zhang},
  \citenamefont {Goyal}, \citenamefont {Gao}, \citenamefont {Sanders},\ and\
  \citenamefont {Simon}}]{Zhang2016}%
  \BibitemOpen
  \bibfield  {author} {\bibinfo {author} {\bibfnamefont {W.-W.}\ \bibnamefont
  {Zhang}}, \bibinfo {author} {\bibfnamefont {S.~K.}\ \bibnamefont {Goyal}},
  \bibinfo {author} {\bibfnamefont {F.}~\bibnamefont {Gao}}, \bibinfo {author}
  {\bibfnamefont {B.~C.}\ \bibnamefont {Sanders}}, \ and\ \bibinfo {author}
  {\bibfnamefont {C.}~\bibnamefont {Simon}},\ }\href {\doibase
  10.1088/1367-2630/18/9/093025} {\bibfield  {journal} {\bibinfo  {journal}
  {New Journal of Physics}\ }\textbf {\bibinfo {volume} {18}},\ \bibinfo
  {pages} {093025} (\bibinfo {year} {2016})}\BibitemShut {NoStop}%
\bibitem [{\citenamefont {Flurin}\ \emph {et~al.}(2017)\citenamefont {Flurin},
  \citenamefont {Ramasesh}, \citenamefont {Hacohen-Gourgy}, \citenamefont
  {Martin}, \citenamefont {Yao},\ and\ \citenamefont {Siddiqi}}]{Catstates2}%
  \BibitemOpen
  \bibfield  {author} {\bibinfo {author} {\bibfnamefont {E.}~\bibnamefont
  {Flurin}}, \bibinfo {author} {\bibfnamefont {V.~V.}\ \bibnamefont
  {Ramasesh}}, \bibinfo {author} {\bibfnamefont {S.}~\bibnamefont
  {Hacohen-Gourgy}}, \bibinfo {author} {\bibfnamefont {L.~S.}\ \bibnamefont
  {Martin}}, \bibinfo {author} {\bibfnamefont {N.~Y.}\ \bibnamefont {Yao}}, \
  and\ \bibinfo {author} {\bibfnamefont {I.}~\bibnamefont {Siddiqi}},\ }\href
  {\doibase 10.1103/PhysRevX.7.031023} {\bibfield  {journal} {\bibinfo
  {journal} {Phys. Rev. X}\ }\textbf {\bibinfo {volume} {7}},\ \bibinfo {pages}
  {031023} (\bibinfo {year} {2017})}\BibitemShut {NoStop}%
\bibitem [{\citenamefont {Giordani}\ \emph {et~al.}(2019)\citenamefont
  {Giordani}, \citenamefont {Polino}, \citenamefont {Emiliani}, \citenamefont
  {Suprano}, \citenamefont {Innocenti}, \citenamefont {Majury}, \citenamefont
  {Marrucci}, \citenamefont {Paternostro}, \citenamefont {Ferraro},
  \citenamefont {Spagnolo},\ and\ \citenamefont {Sciarrino}}]{Sciarrino2019}%
  \BibitemOpen
  \bibfield  {author} {\bibinfo {author} {\bibfnamefont {T.}~\bibnamefont
  {Giordani}}, \bibinfo {author} {\bibfnamefont {E.}~\bibnamefont {Polino}},
  \bibinfo {author} {\bibfnamefont {S.}~\bibnamefont {Emiliani}}, \bibinfo
  {author} {\bibfnamefont {A.}~\bibnamefont {Suprano}}, \bibinfo {author}
  {\bibfnamefont {L.}~\bibnamefont {Innocenti}}, \bibinfo {author}
  {\bibfnamefont {H.}~\bibnamefont {Majury}}, \bibinfo {author} {\bibfnamefont
  {L.}~\bibnamefont {Marrucci}}, \bibinfo {author} {\bibfnamefont
  {M.}~\bibnamefont {Paternostro}}, \bibinfo {author} {\bibfnamefont
  {A.}~\bibnamefont {Ferraro}}, \bibinfo {author} {\bibfnamefont
  {N.}~\bibnamefont {Spagnolo}}, \ and\ \bibinfo {author} {\bibfnamefont
  {F.}~\bibnamefont {Sciarrino}},\ }\href {\doibase
  10.1103/PhysRevLett.122.020503} {\bibfield  {journal} {\bibinfo  {journal}
  {Phys. Rev. Lett.}\ }\textbf {\bibinfo {volume} {122}},\ \bibinfo {pages}
  {020503} (\bibinfo {year} {2019})}\BibitemShut {NoStop}%
\bibitem [{\citenamefont {Vieira}\ \emph {et~al.}(2013)\citenamefont {Vieira},
  \citenamefont {Amorim},\ and\ \citenamefont {Rigolin}}]{Rigolin2013}%
  \BibitemOpen
  \bibfield  {author} {\bibinfo {author} {\bibfnamefont {R.}~\bibnamefont
  {Vieira}}, \bibinfo {author} {\bibfnamefont {E.~P.~M.}\ \bibnamefont
  {Amorim}}, \ and\ \bibinfo {author} {\bibfnamefont {G.}~\bibnamefont
  {Rigolin}},\ }\href {\doibase 10.1103/PhysRevLett.111.180503} {\bibfield
  {journal} {\bibinfo  {journal} {Phys. Rev. Lett.}\ }\textbf {\bibinfo
  {volume} {111}},\ \bibinfo {pages} {180503} (\bibinfo {year}
  {2013})}\BibitemShut {NoStop}%
\bibitem [{\citenamefont {Vieira}\ \emph {et~al.}(2014)\citenamefont {Vieira},
  \citenamefont {Amorim},\ and\ \citenamefont {Rigolin}}]{Rigolin2014}%
  \BibitemOpen
  \bibfield  {author} {\bibinfo {author} {\bibfnamefont {R.}~\bibnamefont
  {Vieira}}, \bibinfo {author} {\bibfnamefont {E.~P.~M.}\ \bibnamefont
  {Amorim}}, \ and\ \bibinfo {author} {\bibfnamefont {G.}~\bibnamefont
  {Rigolin}},\ }\href {\doibase 10.1103/PhysRevA.89.042307} {\bibfield
  {journal} {\bibinfo  {journal} {Phys. Rev. A}\ }\textbf {\bibinfo {volume}
  {89}},\ \bibinfo {pages} {042307} (\bibinfo {year} {2014})}\BibitemShut
  {NoStop}%
\bibitem [{\citenamefont {Abal}\ \emph {et~al.}(2006)\citenamefont {Abal},
  \citenamefont {Siri}, \citenamefont {Romanelli},\ and\ \citenamefont
  {Donangelo}}]{Abal2006}%
  \BibitemOpen
  \bibfield  {author} {\bibinfo {author} {\bibfnamefont {G.}~\bibnamefont
  {Abal}}, \bibinfo {author} {\bibfnamefont {R.}~\bibnamefont {Siri}}, \bibinfo
  {author} {\bibfnamefont {A.}~\bibnamefont {Romanelli}}, \ and\ \bibinfo
  {author} {\bibfnamefont {R.}~\bibnamefont {Donangelo}},\ }\href {\doibase
  10.1103/PhysRevA.73.042302} {\bibfield  {journal} {\bibinfo  {journal} {Phys.
  Rev. A}\ }\textbf {\bibinfo {volume} {73}},\ \bibinfo {pages} {042302}
  (\bibinfo {year} {2006})}\BibitemShut {NoStop}%
\bibitem [{\citenamefont {Orthey}\ and\ \citenamefont
  {Amorim}(2017)}]{Orthey2017}%
  \BibitemOpen
  \bibfield  {author} {\bibinfo {author} {\bibfnamefont {A.~C.}\ \bibnamefont
  {Orthey}}\ and\ \bibinfo {author} {\bibfnamefont {E.~P.~M.}\ \bibnamefont
  {Amorim}},\ }\href {\doibase 10.1007/s11128-017-1672-1} {\bibfield  {journal}
  {\bibinfo  {journal} {Quantum Inf. Process.}\ }\textbf {\bibinfo {volume}
  {16}},\ \bibinfo {pages} {224} (\bibinfo {year} {2017})}\BibitemShut
  {NoStop}%
\bibitem [{\citenamefont {Wang}\ \emph {et~al.}(2018)\citenamefont {Wang},
  \citenamefont {Xu}, \citenamefont {Pan}, \citenamefont {Sun}, \citenamefont
  {Xu}, \citenamefont {Chen}, \citenamefont {Han}, \citenamefont {Li},\ and\
  \citenamefont {Guo}}]{Wang:18}%
  \BibitemOpen
  \bibfield  {author} {\bibinfo {author} {\bibfnamefont {Q.-Q.}\ \bibnamefont
  {Wang}}, \bibinfo {author} {\bibfnamefont {X.-Y.}\ \bibnamefont {Xu}},
  \bibinfo {author} {\bibfnamefont {W.-W.}\ \bibnamefont {Pan}}, \bibinfo
  {author} {\bibfnamefont {K.}~\bibnamefont {Sun}}, \bibinfo {author}
  {\bibfnamefont {J.-S.}\ \bibnamefont {Xu}}, \bibinfo {author} {\bibfnamefont
  {G.}~\bibnamefont {Chen}}, \bibinfo {author} {\bibfnamefont {Y.-J.}\
  \bibnamefont {Han}}, \bibinfo {author} {\bibfnamefont {C.-F.}\ \bibnamefont
  {Li}}, \ and\ \bibinfo {author} {\bibfnamefont {G.-C.}\ \bibnamefont {Guo}},\
  }\href {\doibase 10.1364/OPTICA.5.001136} {\bibfield  {journal} {\bibinfo
  {journal} {Optica}\ }\textbf {\bibinfo {volume} {5}},\ \bibinfo {pages}
  {1136} (\bibinfo {year} {2018})}\BibitemShut {NoStop}%
\bibitem [{\citenamefont {Chandrashekar}\ \emph {et~al.}(2008)\citenamefont
  {Chandrashekar}, \citenamefont {Srikanth},\ and\ \citenamefont
  {Laflamme}}]{Chandrashekar2008}%
  \BibitemOpen
  \bibfield  {author} {\bibinfo {author} {\bibfnamefont {C.~M.}\ \bibnamefont
  {Chandrashekar}}, \bibinfo {author} {\bibfnamefont {R.}~\bibnamefont
  {Srikanth}}, \ and\ \bibinfo {author} {\bibfnamefont {R.}~\bibnamefont
  {Laflamme}},\ }\href {\doibase 10.1103/PhysRevA.77.032326} {\bibfield
  {journal} {\bibinfo  {journal} {Phys. Rev. A}\ }\textbf {\bibinfo {volume}
  {77}},\ \bibinfo {pages} {032326} (\bibinfo {year} {2008})}\BibitemShut
  {NoStop}%
\bibitem [{\citenamefont {Reuvers}(2017)}]{Reuvers2018}%
  \BibitemOpen
  \bibfield  {author} {\bibinfo {author} {\bibfnamefont {R.}~\bibnamefont
  {Reuvers}},\ }\href
  {https://royalsocietypublishing.org/doi/full/10.1098/rspa.2018.0023}
  {\bibfield  {journal} {\bibinfo  {journal} {Proc. Royal Soc. Lond. A}\
  }\textbf {\bibinfo {volume} {474}},\ \bibinfo {pages} {20180023} (\bibinfo
  {year} {2017})}\BibitemShut {NoStop}%
\bibitem [{\citenamefont {Zeng}\ and\ \citenamefont {Yong}(2017)}]{Zeng2017}%
  \BibitemOpen
  \bibfield  {author} {\bibinfo {author} {\bibfnamefont {M.}~\bibnamefont
  {Zeng}}\ and\ \bibinfo {author} {\bibfnamefont {E.~H.}\ \bibnamefont
  {Yong}},\ }\href {\doibase 10.1038/s41598-017-12077-0} {\bibfield  {journal}
  {\bibinfo  {journal} {Sci. Rep.}\ }\textbf {\bibinfo {volume} {7}},\ \bibinfo
  {pages} {12024} (\bibinfo {year} {2017})}\BibitemShut {NoStop}%
\bibitem [{\citenamefont {Wales}\ and\ \citenamefont
  {Doye}(1997)}]{wales_1997}%
  \BibitemOpen
  \bibfield  {author} {\bibinfo {author} {\bibfnamefont {D.~J.}\ \bibnamefont
  {Wales}}\ and\ \bibinfo {author} {\bibfnamefont {J.~P.~K.}\ \bibnamefont
  {Doye}},\ }\href {https://doi.org/10.1021/jp970984n} {\bibfield  {journal}
  {\bibinfo  {journal} {J Phys Chem A}\ }\textbf {\bibinfo {volume} {101}},\
  \bibinfo {pages} {5111} (\bibinfo {year} {1997})}\BibitemShut {NoStop}%
\bibitem [{\citenamefont {Cardano}\ \emph {et~al.}(2016)\citenamefont
  {Cardano}, \citenamefont {Maffei}, \citenamefont {Massa}, \citenamefont
  {Piccirillo}, \citenamefont {de~Lisio}, \citenamefont {De~Filippis},
  \citenamefont {Cataudella}, \citenamefont {Santamato},\ and\ \citenamefont
  {Marrucci}}]{Cardano2016}%
  \BibitemOpen
  \bibfield  {author} {\bibinfo {author} {\bibfnamefont {F.}~\bibnamefont
  {Cardano}}, \bibinfo {author} {\bibfnamefont {M.}~\bibnamefont {Maffei}},
  \bibinfo {author} {\bibfnamefont {F.}~\bibnamefont {Massa}}, \bibinfo
  {author} {\bibfnamefont {B.}~\bibnamefont {Piccirillo}}, \bibinfo {author}
  {\bibfnamefont {C.}~\bibnamefont {de~Lisio}}, \bibinfo {author}
  {\bibfnamefont {G.}~\bibnamefont {De~Filippis}}, \bibinfo {author}
  {\bibfnamefont {V.}~\bibnamefont {Cataudella}}, \bibinfo {author}
  {\bibfnamefont {E.}~\bibnamefont {Santamato}}, \ and\ \bibinfo {author}
  {\bibfnamefont {L.}~\bibnamefont {Marrucci}},\ }\href {\doibase
  10.1038/ncomms11439} {\bibfield  {journal} {\bibinfo  {journal} {Nat. Comm.}\
  }\textbf {\bibinfo {volume} {7}},\ \bibinfo {pages} {11439} (\bibinfo {year}
  {2016})}\BibitemShut {NoStop}%
\bibitem [{\citenamefont {{D'Errico}}\ \emph {et~al.}(2018)\citenamefont
  {{D'Errico}}, \citenamefont {{Cardano}}, \citenamefont {{Maffei}},
  \citenamefont {{Dauphin}}, \citenamefont {{Barboza}}, \citenamefont
  {{Esposito}}, \citenamefont {{Piccirillo}}, \citenamefont {{Lewenstein}},
  \citenamefont {{Massignan}},\ and\ \citenamefont {{Marrucci}}}]{Derrico2018}%
  \BibitemOpen
  \bibfield  {author} {\bibinfo {author} {\bibfnamefont {A.}~\bibnamefont
  {{D'Errico}}}, \bibinfo {author} {\bibfnamefont {F.}~\bibnamefont
  {{Cardano}}}, \bibinfo {author} {\bibfnamefont {M.}~\bibnamefont {{Maffei}}},
  \bibinfo {author} {\bibfnamefont {A.}~\bibnamefont {{Dauphin}}}, \bibinfo
  {author} {\bibfnamefont {R.}~\bibnamefont {{Barboza}}}, \bibinfo {author}
  {\bibfnamefont {C.}~\bibnamefont {{Esposito}}}, \bibinfo {author}
  {\bibfnamefont {B.}~\bibnamefont {{Piccirillo}}}, \bibinfo {author}
  {\bibfnamefont {M.}~\bibnamefont {{Lewenstein}}}, \bibinfo {author}
  {\bibfnamefont {P.}~\bibnamefont {{Massignan}}}, \ and\ \bibinfo {author}
  {\bibfnamefont {L.}~\bibnamefont {{Marrucci}}},\ }\href@noop {} {\bibfield
  {journal} {\bibinfo  {journal} {arXiv e-prints}\ ,\ \bibinfo {eid}
  {arXiv:1811.04001}} (\bibinfo {year} {2018})},\ \Eprint
  {http://arxiv.org/abs/1811.04001} {1811.04001} \BibitemShut {NoStop}%
\bibitem [{\citenamefont {Luo}\ \emph {et~al.}(2019)\citenamefont {Luo},
  \citenamefont {Zhong}, \citenamefont {Erhard}, \citenamefont {Wang},
  \citenamefont {Peng}, \citenamefont {Krenn}, \citenamefont {Jiang},
  \citenamefont {Li}, \citenamefont {Liu}, \citenamefont {Lu}, \citenamefont
  {Zeilinger},\ and\ \citenamefont {Pan}}]{Pan2019}%
  \BibitemOpen
  \bibfield  {author} {\bibinfo {author} {\bibfnamefont {Y.-H.}\ \bibnamefont
  {Luo}}, \bibinfo {author} {\bibfnamefont {H.-S.}\ \bibnamefont {Zhong}},
  \bibinfo {author} {\bibfnamefont {M.}~\bibnamefont {Erhard}}, \bibinfo
  {author} {\bibfnamefont {X.-L.}\ \bibnamefont {Wang}}, \bibinfo {author}
  {\bibfnamefont {L.-C.}\ \bibnamefont {Peng}}, \bibinfo {author}
  {\bibfnamefont {M.}~\bibnamefont {Krenn}}, \bibinfo {author} {\bibfnamefont
  {X.}~\bibnamefont {Jiang}}, \bibinfo {author} {\bibfnamefont
  {L.}~\bibnamefont {Li}}, \bibinfo {author} {\bibfnamefont {N.-L.}\
  \bibnamefont {Liu}}, \bibinfo {author} {\bibfnamefont {C.-Y.}\ \bibnamefont
  {Lu}}, \bibinfo {author} {\bibfnamefont {A.}~\bibnamefont {Zeilinger}}, \
  and\ \bibinfo {author} {\bibfnamefont {J.-W.}\ \bibnamefont {Pan}},\ }\href
  {\doibase 10.1103/PhysRevLett.123.070505} {\bibfield  {journal} {\bibinfo
  {journal} {Phys. Rev. Lett.}\ }\textbf {\bibinfo {volume} {123}},\ \bibinfo
  {pages} {070505} (\bibinfo {year} {2019})}\BibitemShut {NoStop}%
\end{thebibliography}
\end{document}